\documentclass[a4paper,11pt]{article}
\usepackage{mathtools}
\usepackage{pos}
\usepackage{textcomp}
\usepackage{lipsum} 
\usepackage{gensymb}
\usepackage[left]{lineno}

\title{Search for High Energy Neutrinos from Infrared Flares}

\ShortTitle{Neutrinos and flares}

\author{The IceCube Collaboration \\{\normalsize \normalfont(a complete list of authors can be found at the end of the proceedings)}\\}

\emailAdd{teresa.pernice@desy.de}
\emailAdd{sommani@astro.ruhr-uni-bochum.de}

\abstract{

IceCube has detected a diffuse flux of high-energy neutrinos, whose origin still remains uncertain. Accreting supermassive black holes (SMBHs) have been proposed as plausible sources of neutrinos. Candidate sources include AT2019dsg, which is likely a stellar tidal disruption event (TDE), and AT2019dfr, an AGN flare.
Both present delayed emission in the IR band with respect to the optical signal. This emission can be interpreted as the reprocessing of X-rays to optical light of the flare by dust located in a torus around the SMBH. An additional study using an optically detected sample of 63 accretion flares revealed another candidate as a potential high-energy neutrino counterpart: AT2019aalc, which is also accompanied by a dust echo. However, follow-up stacking analysis of the 63 nuclear flares using the full IceCube data sample did not show any significant excess over background. Motivated by these three suggested neutrino-TDE correlations, we analyze a more extensive catalog of IR flares, 823 dust-echo-like flares identified using WISE satellite data, against the IceCube 10-year sample of track events from the Northern Sky. Our analysis aims to perform sensitivity studies and assess the potential detectability of neutrino emission from these types of accretion flares. In addition, we carry out a correlation study of the 823 dust echo-like flares against a revised catalog of IceCube high-purity astrophysical alerts, and reevaluate the previous study of 63 nuclear flares against the same revised alerts sample.

\vspace{4mm}

{\bfseries Corresponding authors:}

Teresa Pernice$^{1}$, 
Giacomo Sommani$^{2*}$,\\

{$^{1}$ \itshape DESY}\\
{$^{2}$ \itshape Ruhr-Universität Bochum}\\[4mm]
$^*$ Presenter
}

\ConferenceLogo{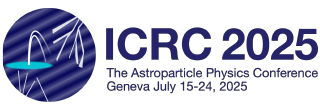}

\FullConference{39th International Cosmic Ray Conference (ICRC2025)\\
 15–24 July 2025\\
Geneva, Switzerland\\}

\begin{document}

\maketitle

\vspace{-3mm}
\section{Introduction}\label{sec1}
\vspace{-3mm}

IceCube is a neutrino detector located at the South Pole, consisting of 1 km$^3$ of instrumented glacial ice.
The detector uses 5,160 Digital Optical Modules (DOMs) as its detection units, deployed along 86 vertical strings~\cite{Aartsen:2016nxy}.
When a neutrino enters the ice, it interacts with the transparent medium, producing secondary particles, among which are muons that can travel for kilometers through the detector.
All the charged secondary particles induce Cherenkov light, which is then detected by the DOMs.
Cherenkov light primarily produced along a muon track characterizes so-called track events.

Over the course of its operation, IceCube has detected a diffuse flux of high-energy neutrinos \cite{PhysRevLett.113.101101}, with the only two high-confidence extragalactic sources identified to date being the accreting supermassive black holes (SMBHs) TXS~0506+056 \cite{TXS2018} and NGC~1068 \cite{NGC1068}.
To investigate the possibility of neutrinos originating from powerful transient events, IceCube implemented a realtime system in 2016~\cite{Aartsen:2016lmt}.
As soon as a track event with a high probability of an astrophysical signal is detected, an alert is issued, permitting prompt follow-up from other instruments.
Three of these track alerts have been associated with time-variable accretion events on supermassive black holes (SMBHs)~\cite{Stein2021, Reusch22, van_Velzen_2024}.
These flares, known as Tidal Disruption Events (TDEs), are rare astrophysical transient events that occur when a star passes close to a SMBH, that is believed to reside in the center of almost every galaxy.
The star is tidally disrupted, and part of the debris forms an accretion disk around the black hole.
The infalling material emits a bright flare across the electromagnetic spectrum.
If there is dust surrounding the SMBH, this radiation heats the dust grains, which then re-emit in the infrared (IR) wavelength range.
This delayed IR emission, also called dust echoes, has been observed in all three coincident events~\cite{van_Velzen_2024}. Motivated by these correlations, a stacking analysis was first performed on a sample of 63 nuclear flares without finding any significant excess over background. However, an upper limit for harder ($\gamma < 2 $) spectral index was reported, for the detection of three neutrino alert events (IC-191001A, IC-191119A, and IC-200530A) from this type of source \cite{necker2023searchhighenergyneutrinostdelike}. To enable a more sensitive search for neutrinos from accretion flares with bright dust echoes, the Flaires catalog was compiled \cite{Necker_2025}. Starting from a sample of $O(10^6)$ light curves, \cite{Necker_2025} identified dust echoes of possible accretion events using data collected by the WISE satellite and compiled a catalog of 823 dust-echo-like flares. Informed by the possibility of delayed IR emission occuring up to one year after neutrino detection \cite{Winter_2023}, we first employ a time-dependent stacking analysis framework using the Flaires catalog and the 10-year dataset of muon tracks from the Northern Sky, which benefits from superior angular resolution and reduced atmospheric background. We will refer to this dataset as \textit{Northern Tracks}.
Moreover, the reconstruction used in the track alerts originally associated with the TDEs was updated in September 2024~\cite{gcn_update} and led to the updated catalog of track alerts, IceCat-2, where the three TDEs originally coincident are no longer inside the new contours~\cite{icecat2}.
This motivates two other stacking analyses that will test separately both Flaires and the 63 accretion flares in the nuclear sample with IceCat-2. 

The description of the stacking framework can be found in~\textsection\ref{stackings}, where we discuss the different approaches adopted in this analysis: correlation with full archival data in~\textsection\ref{subsec:flairestack}, correlation with neutrino alerts in ~\textsection\ref{subsec:alertstack}. The sensitivity results will be presented in Section ~\textsection\ref{sec:sensitivity} for the three analysis and the conclusions and future perspectives can be found in ~\textsection\ref{sec3}.

\vspace{-3mm}
\section{Stacking Analyses}\label{stackings}
\vspace{-3mm}

Two different neutrino datasets are going to be used in the stacking analyses presented here.
The first is Northern Tracks (see~\textsection\ref{sec1}).
This dataset has loose energy cuts, resulting in a background-dominated sample\footnote{Previous versions of the same dataset had an even higher background contamination.}, but allowing for high sensitivities through adequate statistical analysis, particularly below 100 TeV.
The second neutrino dataset is IceCat-2~\cite{icecat2}, which consists of archival track alerts, i.e., selected high-energy tracks, with the most up-to-date reconstruction.
This high-energy selection reduces the catalog size and cancels out the Northern Tracks' sensitivity over 100 GeV to 100 TeV.
However, the background contamination is significantly less, with an average signalness, i.e., probability of astrophysical origin, between 40\% and 50\%.
This dataset of archival alerts is more sensitive to sources with high-energy emission above 100~TeV, and benefits from more accurate reconstruction~\cite{icecat2, realtimereco}.

Northern Tracks is used in the analysis described in~\textsection\ref{subsec:flairestack}, which tests only the Flaires catalog~\cite{Necker_2025}.
IceCat-2 is used in two different analyses introduced in~\textsection\ref{subsec:alertstack}.
These analyses test both the Flaires catalog~\textsection\ref{subsubsec:alertstack_flaires} and also the 63 nuclear flares that brought the first evident correlation between TDEs and IceCube alerts (see~\textsection\ref{sec1})~\textsection\ref{subsubsec:alertstack_sjoert}.
All these analyses are under development and here are reported only preliminary sensitivities~\textsection\ref{sec:sensitivity}.

\subsection{Stacking Analysis with full archival data}\label{subsec:flairestack}

For this analysis, the unbinned likelihood method is used to evaluate the correlation between the sources in the Flaires catalog and neutrinos in the Northern Tracks dataset, and estimate the contribution of these flares to the overall diffuse neutrino flux.
From the Flaires catalog, we select sources with known bolometric fluence and exclude IR-flaring objects that are also identified as synchrotron emitters from jets to remove contamination from known AGNs, quasars, blazars and their radio/X-ray counterparts in order to have a high purity sample of TDEs. After applying these cuts, we are left with a total of 528 IR flares. Of these sources, we select the 394 located in the Northern Sky.

\begin{figure}[h!]
\centering
\begin{minipage}{0.46\linewidth}
    \centering
    \includegraphics[width=\linewidth]{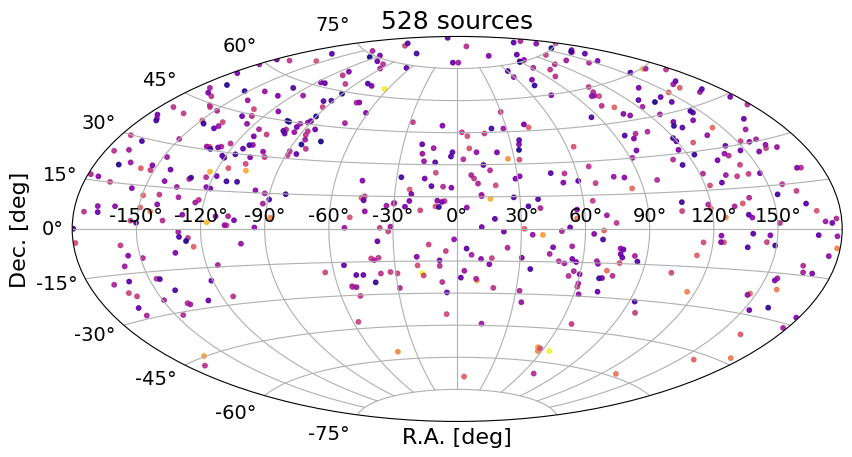}
    \label{fig:northern}
\end{minipage}
\hfill
\begin{minipage}{0.53\linewidth}
    \centering
    \includegraphics[width=\linewidth]{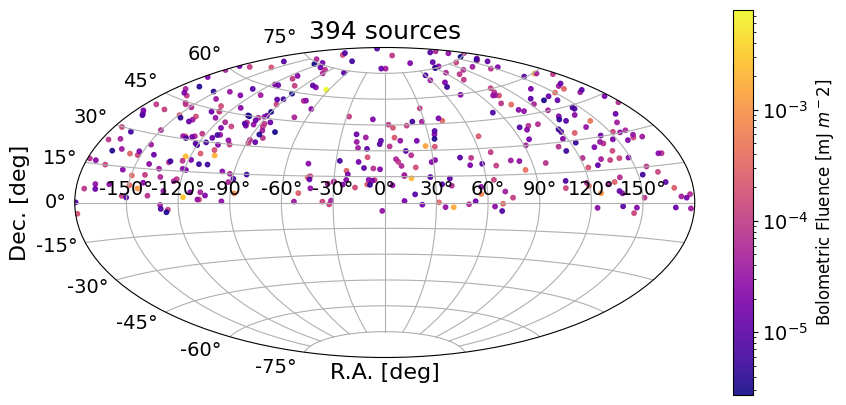}
    \label{fig:southern}
\end{minipage}
\caption{Skymaps of the sources in the Flaires catalog. The left plot shows all 528 sources while the plot on the right shows only the sources in the Northern Sky, which will be used for this analysis using Northern tracks. The colormap shows the total bolometric fluence $mJ\,m^{-2}$ for each source, which is used as weight in the likelihood method.}
\label{fig:skymaps}
\end{figure}

Starting from the standard point source likelihood approach, we use as signal PDF the contribution of the given source population $S = \sum_{j=1}^M w_j S_j(\nu_i, \gamma)$, where M is the number of sources, and $w_j$ is the weight assigned to each source based on its relative contribution to the combined signal. In this study, each source contribution is pre-determined by weighting each catalog's entry by its bolometric fluence (given in the catalog). This quantity was estimated by integrating the bolometric luminosity over the flare duration, using discrete WISE observations \cite{Necker_2025}. The likelihood function then becomes:
\begin{equation}
L(n_s, \gamma) = \prod_{i=1}^{N} \left(\frac{n_s}{N} \prod_{j=1}^{M}w_j S(\nu_{i}, \gamma) + (1 - \frac{n_{s}}{N}) B(\nu_i) \right)\,.
\end{equation}
The first product runs over all the ${\displaystyle N}$ neutrino events in dataset where ${\mathcal {S}}(\nu _{i},\gamma )$ and ${\displaystyle {\mathcal {B}}(\nu _{i})}$ represent, respectively, the signal and the background PDF for the $i_{th}$ neutrino $\nu _{i}$. $n_s$ is the estimated signal strength in the dataset and $\gamma$ is the spectral index of the power law spectrum assumed for the signal PDF. 

Both signal and background PDFs can be factorized into three components: spatial, energy, and temporal PDFs.
The spatial PDF for the signal is modeled using 3D Kernel Densitiy Estimator (KDEs) \cite{NGC1068}. This method  reflects the detector response for a given spectral index and includes dependecies on opening angle, reconstructed energy, and angular uncertainty. The energy component of the signal PDF is generated using Monte Carlo (MC) simulated astrophysical events. A power-law injection model is assumed and to each MC event a weight that reflects the assumed astrophysical spectrum is assigned. Only events within $\pm 5^{\circ}$ in declination of each source are considered. Then, a ($\sin(\delta)$, logE) histogram is made, and then smoothed with 2D splines to represent the gamma-dependent energy PDFs. The temporal signal PDF assumes a 1-year box model starting from the peak of the dust echo to one year before, which is the compatibility time window within which we expect to have a correlated neutrino emission during SMBH accretion events \cite{Winter_2023}.

The background PDF is built from MC simulations of atmospheric background. The background spatial PDF depends only on declination: an histogram in $\sin(\delta)$ is created with atmospheric weights, and smoothed via a spline fitting. Background events are randomly drawn from Monte Carlo simulations, following the atmospheric neutrino background model. Like the signal PDF, events are binned in ($\sin(\delta)$, logE) and weighted by their atmospheric flux weight. For the temporal component, since the dataset is dominated by background, we assume a uniform time distribution.

To evaluate if there is significant evidence for signal, the likelihood ratio between the background-only hypothesis and the signal+background hypothesis is then estimated:
\begin{equation}\label{eq:TS}
    \mathrm{TS} = -2 \log \frac{L(n_s=0)}{L(n_s, \gamma)}\,,
\end{equation}
where $L(n_s = 0) = \sum_{i=1}^{N}B(\nu_i)$. The resulting TS is then compared to a distribution of TS from MC datasets, which are obtained by scrambling the neutrino declinations and keeping fixed the IR flares positions (see ~\textsection\ref{subsubsec:alertstack_flaires}).

\subsection{Stacking Analyses with IceCat-2}\label{subsec:alertstack}

These analyses use a stacking method with a likelihood where the signal hypothesis states that the single neutrino alert originates from a source in the catalog, while the background hypothesis says that it is not correlated with it.
This method for stacking with track alerts was first developed by~\cite{cristinastacking} and makes use of the full probability maps for the single neutrino alert taken from IceCat-2~\cite{icecat2}.
These probability maps consist of full-sky maps where a probability for the neutrino to originate from every possible direction is assigned.
The likelihood~$L$ is the weighted sum of a signal PDF, $S$, and a background PDF, $B$,
\begin{equation}\label{eq:general_likelihood}
    L = \frac{n_S}{N}S + \left(1-\frac{n_S}{N}\right)B\,,
\end{equation}
where $n_S$ is the number of signal events and $N$ is the total number of signal and background neutrino candidate events.
The test statistic (TS) for a single neutrino $i$ ($N=1$) becomes
\begin{equation}\label{eq:alert_likelihood}
    \mathrm{TS}_i = -2\log{\left[\frac{L(n_S=0)}{L(n_S=1)}\right]} = -2\log{\frac{B}{\hat{S}}}\,.
\end{equation}
TS$_i$ is evaluated on the source in the catalog that maximizes its value.
If TS$_i<0$, the background hypothesis is chosen, i.e., $S/B=1$.
The signal PDF convolutes three contributions: the alert signalness, a probability from the probability map in the direction of the source, and a weight assigned based on the assumed neutrino production mechanism and time correlation.
The weights are different in the two analyses~\textsection\ref{subsubsec:alertstack_sjoert} and~\textsection\ref{subsubsec:alertstack_flaires} and comprise also two different compatibility time windows.

The background PDF describes the probability of finding the source $b$ by chance in its position given the tested catalog.
If the sources in the catalog are distributed isotropically (which is not the case in~\textsection\ref{subsubsec:alertstack_sjoert} and~\textsection\ref{subsubsec:alertstack_flaires}), $B=1/4\pi$.
Finally, the total TS is calculated as $\mathrm{TS}=\sum_i{\mathrm{TS}_i}$.

Under this framework, to perform sensitivity studies~(see~\textsection\ref{sec:sensitivity}), the position of the flares is randomly generated, while the neutrinos are fixed, differently from~\textsection\ref{subsec:flairestack}.
For the signal-injection, a single flare is randomly selected according to its weight $w_b$, and its peak time is artificially changed to be within the desired compatibility time window~(which depends on the specific analysis) with the neutrino event.
Its position in the sky is also chosen randomly, following the probability map of the single neutrino alert.
If more signal neutrinos are injected, the same flares are not counted more than once.

This method of stacking neutrino alerts is applied to two catalogs of infrared flares: the original 63 nuclear flares~\cite{van_Velzen_2024}~\textsection\ref{subsubsec:alertstack_sjoert}, and the Flaires catalog~\cite{Necker_2025}~\textsection\ref{subsubsec:alertstack_flaires}.

\subsubsection{Alert stacking with 63 nuclear flares}\label{subsubsec:alertstack_sjoert}

To implement the stacking method explained in~\textsection\ref{subsec:alertstack}, a single weight $w_b$ for the single accretion flare must be chosen to be used in the signal PDF $S$, and the background PDF $B$ must be properly expressed.

The test statistic used in~\cite{van_Velzen_2024} is of the form
\begin{equation}\label{eq:sjoert_ts}
    \mathrm{TS} = 2\log{\frac{\hat{L}_S}{L_0}} = 2\log{\left(\prod_i{S_{\mathrm{signalness},\,\nu}\,\cdot\,N_{\mathrm{chance},\,\nu}\,\cdot\,R_b}\right)}\,,
\end{equation}
where $i$ is the index of the $i$-th alert, $N_{\mathrm{chance},\,\nu}$ is number of chance coincidences that are expected for that single neutrino alert (taking into account the amount of flares in the catalog and the size of the $i$-th alert), and $R_b$ is a ratio between the probabilities of observing precise features of the single flare under the signal and background hypotheses~\cite{van_Velzen_2024}.
As weight to the 63 accretion flares, we take the same $R_b$ used in~\cite{van_Velzen_2024}\footnote{The code of the work was made available by the authors in \url{https://zenodo.org/records/7026636}.}.
Moreover, in~\cite{van_Velzen_2024} a compatibility time window of 1~year after the optical peak of the accretion flare is adopted (in~\textsection\ref{subsec:flairestack} and~\textsection\ref{subsubsec:alertstack_flaires} the time window is before the IR peak instead, which typically occurs order of 100 days after the optical peak).
Here, we follow the same approach by setting $w_b$ to zero if the neutrino is outside the same time window.
As a consequence, we do not consider the neutrino events that can not fall within any time window of the accretion flares.
After this cut, we are left with 73 neutrino alerts (from the original 365~\cite{icecat2}).

For the background PDF $B$, we assumed an isotropic distribution of the 63 accretion flares in the sky to evaluate some preliminary sensitivities.
A more accurate approach, which is not yet implemented but is a work in progress, would take into account the non-uniform source distribution by dividing the sky into pixels and estimating the density of sources in each pixel, from which the data-modeled PDF would be extracted.
Nevertheless, the difference between the two approaches is expected to be small.

\subsubsection{Alert stacking with the Flaires catalog}\label{subsubsec:alertstack_flaires}

For the Flaires catalog, the bolometric fluence is taken as weight $w_b$ for the infrared flares, consistently with~\textsection\ref{subsec:flairestack}.
Moreover, the same compatibility time window of one year before the IR peak of~\textsection\ref{subsec:flairestack} is also used in this case (different from the one in~\textsection\ref{subsubsec:alertstack_sjoert}).
We do not consider the neutrino events that can not fall within any time window of the IR flares in this case as well.
After this cut, we are left with 298 neutrino alerts (from the original 365~\cite{icecat2}).

\begin{figure}[h!]
\centering
\includegraphics[width=0.64\linewidth]{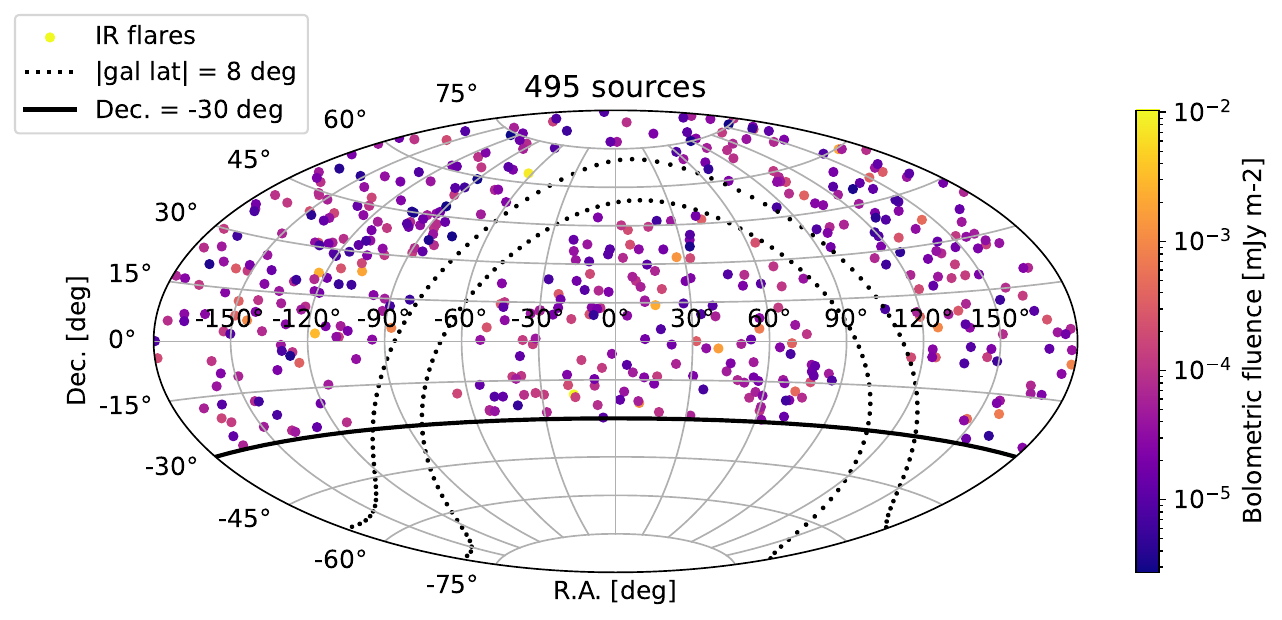}
\caption{
Skymap of the 495 selected sources from the Flaires catalog~\cite{Necker_2025} for the stacking analysis using IceCat-2 neutrino data~\cite{icecat2}.
The colormap shows the total bolometric fluence mJ\,m$^{-2}$ for each source, which is used as a weight in the likelihood method.
}\label{fig:flaires_alertstack}
\end{figure}

Regarding the background PDF $B$, the same problem of non-isotropic distribution of the sources emerges.
As in~\textsection\ref{subsubsec:alertstack_sjoert}, $B$ is preliminarily assumed isotropic with the perspective of updating it to a more realistic description soon.
To make the isotropic assumption more valid in this preliminary approach, we also apply some additional cuts.
We take the same 528 IR flares with known bolometric fluence and without synchrotron emitters as in~\textsection\ref{subsec:flairestack}.
As it is possible to see from the upper panel of Fig.~\ref{fig:skymaps}, within the Galactic plane and below $-30$~deg of declination, the sources are by eye much more sparse than elsewhere.
Therefore, we exclude all IR flares below $-30^\circ$ of declination and within $8^\circ$ from the galactic plane (imitating the scrambling approach used in~\cite{van_Velzen_2024}).
The areas of the sky outside the cuts are not considered, so that, within the remaining portion of the sky, the assumption of isotropy is more valid.
After these cuts, we are left with 495 IR flares~(Fig.~\ref{fig:flaires_alertstack}).

\begin{table}[h]
\centering
\caption{Sensitivity and discovery potential in terms of flux for different spectral indices $\gamma$, only for the analysis with full archival data~\textsection\ref{subsec:flairestack}.}
\begin{tabular}{lccc}
\hline
 & $\gamma=1.0$ & $\gamma=2.0$ & $\gamma=2.5$ \\
\hline
Sensitivity ($\Phi_{0}$ in $\mathrm{GeV}^{-1} \mathrm{cm}^{-2} \mathrm{s}^{-1}$) & $1.32 \times 10^{-15}$ & $5.33 \times 10^{-9}$ & $1.49 \times 10^{-6}$ \\
Discovery Potential ($\Phi_{0}$ in $\mathrm{GeV}^{-1} \mathrm{cm}^{-2} \mathrm{s}^{-1}$) & $1.38 \times 10^{-15}$ & $1.15 \times 10^{-8}$ & $4.08 \times 10^{-6}$ \\
\hline
\end{tabular}
\label{tab:sens_disc}
\end{table}

\begin{figure}[t]
\centering
\includegraphics[width=0.75\linewidth]{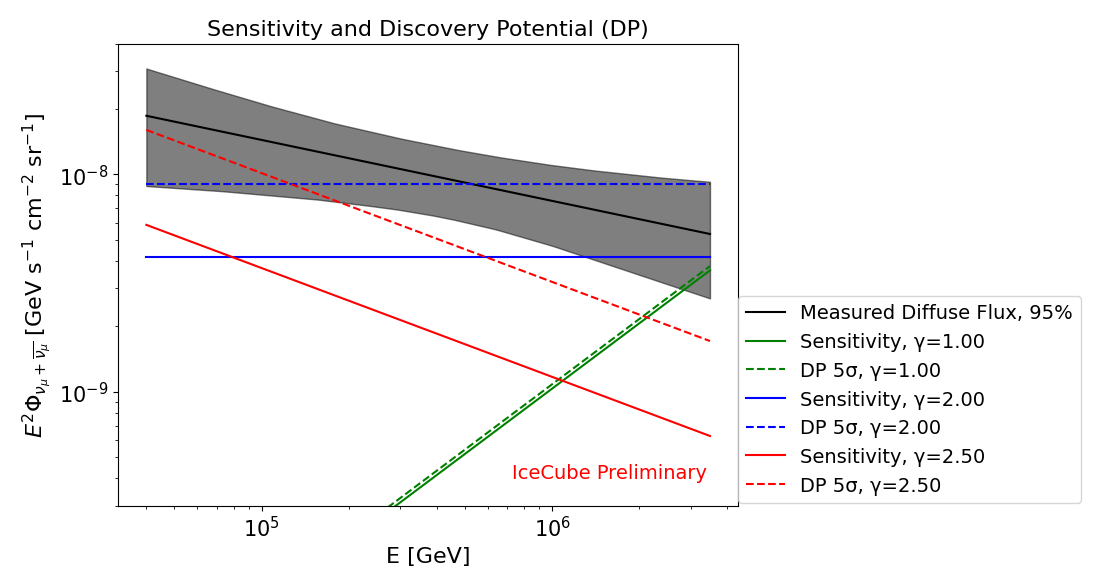}
\caption{
Sensitivity and discovery potential in flux space for the three values of gamma as a function of the energy.
The gray shaded area represents the diffuse flux measured by IceCube using the northern tracks dataset at the 95$\%$ confidence level. 
}\label{fig:combined_sensitivity}
\vspace{-3mm}
\end{figure}

\vspace{-3mm}
\section{Sensitivity studies}\label{sec:sensitivity}
\vspace{-3mm}

To evaluate the sensitivity and the discovery potential of the various stacking analyses, pseudo-experiments are performed by simulating background and signal events as described in the previous section. We then explore the distribution of the test statistic using Eq.~\ref{eq:TS} for the stacking analysis with full archival data (Sec~\textsection\ref{subsec:flairestack}), and $\mathrm{TS}=\sum_i\mathrm{TS}_i$ with $\mathrm{TS_i}$ from Eq.~\ref{eq:alert_likelihood} for the analyses that make use of IceCat-2 (Sec~\textsection\ref{subsec:alertstack}).

The test statistics distributions are then used to compute the sensitivity and the DP of our analyses. The sensitivity determines the injection flux $n_{\mathrm{inj}}$ for which $90\%$ of trials have TS above the background-only median TS value, while the DP defines the injection flux at which $50\%$ of trials exceed the $5 \sigma$ threshold. The DP is extrapolated by fitting a $\chi^2$ to the background TS distribution. The value of the sensitivity and DP in terms of flux are reported in Table~\ref{tab:sens_disc}, while in Fig.~\ref{fig:combined_sensitivity} we show the measured IceCube diffuse flux with northern tracks and the sensitivity and DP values in flux space as a function of energy, for the three values of gamma tested.

For the full archival data, we tested three different values of gamma for the signal hypothesis: $\gamma = 1, 2, 2.5$. The value of $\gamma = 1$ is motivated by results presented in \cite{Winter_2023} for the infrared model.
Regarding the analyses with IceCube track alerts, the signal hypothesis does not depend on an assumed spectral index.
For these analyses, we report only the sensitivity.
The value of the sensitivities and discovery potentials in terms of astrophysical neutrinos can be found in Table~\ref{tab:sens_disc_nu}. The analyses with full archival data and track alerts consider neutrinos in very different energy ranges. Track alerts can be roughly seen as an high-energy selection above $\sim100$~TeV. To provide a more direct comparison between the two neutrino datasets, we include for the full archival data also the estimates that represent the expected number of signal neutrinos above 100~TeV.

\begin{table}[h]
\centering
\caption{
Sensitivity and discovery potential in terms of injected neutrinos for the various analyses data~\textsection\ref{subsec:flairestack}. On the left, the number of signal neutrinos corresponds to the full energy range of the MC data. The values in brackets indicate the expected number of neutrinos above 100 TeV. For the analyses with the track alerts, only the sensitivities are available.
To note that the 2 neutrinos of sensitivity for track alerts with Flaires entail a major improvement in sensitivity in terms of percentage of astrophysical neutrino flux compared to using the 63 accretion flares, as with the Flaires catalog, many more neutrino alerts are used~\textsection\ref{subsubsec:alertstack_sjoert}\textsection\ref{subsubsec:alertstack_flaires}.
}
\begin{tabular}{lccccc}
\hline
 & \multicolumn{3}{c}{Full archival data} & \multicolumn{2}{c}{Neutrino track alerts} \\
 & $\gamma=1.0$ & $\gamma=2.0$ & $\gamma=2.5$ & 63 accretion flares & Flaires\\
\hline
\hline
Sensitivity & 3 (3) & 9 (3) & 20 (1) & 2 & 2 \\
Discovery Potential & 3 (3) & 19 (6) & 56 (4) & - & - \\
\hline
\end{tabular}
\label{tab:sens_disc_nu}
\vspace{-3mm}
\end{table}

\vspace{-3mm}
\section{Conclusion}\label{sec3}
\vspace{-3mm}
We planned several stacking analyses to investigate the correlation between IceCube data and TDEs and reported their sensitivities.
One study made use of the 10 years IceCube northern tracks data sample and a catalog of almost 400 dust-echo-like flares with three values of gamma.
This analysis showed an improvement in sensitivity for a harder source spectrum, which contributes to the measured diffuse flux only at the very highest energies.

The remaining studies will use the IceCube track alerts~\cite{icecat2}, as correlations were hypothesized  between some alerts and TDEs.
One of them will use the same catalog of accretion flares that individuated an excess~\textsection\ref{subsubsec:alertstack_sjoert}\cite{van_Velzen_2024}, and the second one will use the Flaires catalog to have an exact comparison with the approach with full archival data~\textsection\ref{subsec:flairestack}.

Moving forward, we will apply these methods to real IceCube data and shed more light on the contribution of TDEs to the observed diffuse neutrino flux.

\vspace{-3mm}
\begingroup
\footnotesize
\bibliographystyle{ICRC}
\setlength{\bibsep}{1pt}
\bibliography{references}

\providecommand{\href}[2]{#2}\begingroup\raggedright\begin{thebibliography}{10}

\bibitem{Aartsen:2016nxy}
{\bfseries IceCube} Collaboration, M.~G. Aartsen {\em et~al.}, \href{http://dx.doi.org/10.1088/1748-0221/12/03/P03012}{{\em JINST} {\bfseries 12} no.~03, (2017) P03012}.

\bibitem{PhysRevLett.113.101101}
{\bfseries IceCube} Collaboration, M.~G. Aartsen {\em et~al.}, \href{http://dx.doi.org/10.1103/PhysRevLett.113.101101}{{\em Phys. Rev. Lett.} {\bfseries 113} (Sep, 2014) 101101}.

\bibitem{TXS2018}
{\bfseries IceCube} Collaboration, M.~G. Aartsen {\em et~al.}, \href{http://dx.doi.org/10.1126/science.aat1378}{{\em Science} {\bfseries 361} no.~6398, (July, 2018) }.

\bibitem{NGC1068}
{\bfseries IceCube} Collaboration, R.~"Abbasi and others", \href{http://dx.doi.org/10.1126/science.abg3395}{{\em Science} {\bfseries 378} no.~6619, (2022) 538--543}.

\bibitem{Aartsen:2016lmt}
{\bfseries IceCube} Collaboration, M.~G. Aartsen {\em et~al.}, \href{http://dx.doi.org/10.1016/j.astropartphys.2017.05.002}{{\em Astropart. Phys.} {\bfseries 92} (2017) 30--41}.

\bibitem{Stein2021}
R.~Stein {\em et~al.}, \href{http://dx.doi.org/10.1038/s41550-020-01295-8}{{\em Nat. Astron.} {\bfseries 5} no.~5, (2021) 510–518}.

\bibitem{Reusch22}
S.~Reusch {\em et~al.}, \href{http://dx.doi.org/10.1103/PhysRevLett.128.221101}{{\em Phys. Rev. Lett.} {\bfseries 128} (2022) 221101}.

\bibitem{van_Velzen_2024}
S.~van Velzen {\em et~al.}, \href{http://dx.doi.org/10.1093/mnras/stae610}{{\em MNRAS} {\bfseries 529} no.~3, (2024) 2559–2576}.

\bibitem{necker2023searchhighenergyneutrinostdelike}
{\bfseries IceCube} Collaboration, J.~Necker, \href{http://dx.doi.org/10.22323/1.444.1478}{{\em PoS} {\bfseries ICRC2023} (2023) 1478}.

\bibitem{Necker_2025}
J.~Necker {\em et~al.}, \href{http://dx.doi.org/10.1051/0004-6361/202451340}{{\em A\&A} {\bfseries 695} (2025) A228}.

\bibitem{Winter_2023}
W.~Winter and C.~Lunardini, \href{http://dx.doi.org/10.3847/1538-4357/acbe9e}{{\em ApJ} {\bfseries 948} no.~1, (2023) 42}.

\bibitem{gcn_update}
{\bfseries IceCube} Collaboration, {\em GCN} {\bfseries 38267} (2024) .

\bibitem{icecat2}
{\bfseries IceCube} Collaboration, A.~Zegarelli, A.~Franckowiak, G.~Sommani, N.~Valtonen-Mattila, and T.~Yuan, {\em PoS} {\bfseries ICRC2025} (2025) 1224.

\bibitem{realtimereco}
{\bfseries IceCube} Collaboration, G.~Sommani and T.~Yuan, {\em PoS} {\bfseries ICRC2025} (2025) 1184.

\bibitem{cristinastacking}
{\bfseries IceCube} Collaboration, R.~Abbasi {\em et~al.}, \href{http://dx.doi.org/10.3847/1538-4357/acdfcb}{{\em ApJ} {\bfseries 954} no.~1, (2023) 75}.

\end{thebibliography}\endgroup
\vspace{-3mm}

\clearpage

\section*{Full Author List: IceCube Collaboration}

\scriptsize
\noindent
R. Abbasi$^{16}$,
M. Ackermann$^{63}$,
J. Adams$^{17}$,
S. K. Agarwalla$^{39,\: {\rm a}}$,
J. A. Aguilar$^{10}$,
M. Ahlers$^{21}$,
J.M. Alameddine$^{22}$,
S. Ali$^{35}$,
N. M. Amin$^{43}$,
K. Andeen$^{41}$
% ...
\\
\\
% ...
$^{10}$ Universit{\'e} Libre de Bruxelles, Science Faculty CP230, B-1050 Brussels, Belgium \\
% ...
$^{16}$ Department of Physics, Loyola University Chicago, Chicago, IL 60660, USA \\
$^{17}$ Dept. of Physics and Astronomy, University of Canterbury, Private Bag 4800, Christchurch, New Zealand \\
% ...
$^{21}$ Niels Bohr Institute, University of Copenhagen, DK-2100 Copenhagen, Denmark \\
$^{22}$ Dept. of Physics, TU Dortmund University, D-44221 Dortmund, Germany \\
% ...
$^{35}$ Dept. of Physics and Astronomy, University of Kansas, Lawrence, KS 66045, USA \\
% ...
$^{39}$ Dept. of Physics and Wisconsin IceCube Particle Astrophysics Center, University of Wisconsin{\textemdash}Madison, Madison, WI 53706, USA \\
% ...
$^{41}$ Department of Physics, Marquette University, Milwaukee, WI 53201, USA \\
% ...
$^{43}$ Bartol Research Institute and Dept. of Physics and Astronomy, University of Delaware, Newark, DE 19716, USA \\
% ...
$^{63}$ Deutsches Elektronen-Synchrotron DESY, Platanenallee 6, D-15738 Zeuthen, Germany \\
$^{\rm a}$ also at Institute of Physics, Sachivalaya Marg, Sainik School Post, Bhubaneswar 751005, India \\
% ...

\subsection*{Acknowledgments}

\noindent
The authors gratefully acknowledge the support from the following agencies and institutions:
USA {\textendash} U.S. National Science Foundation-Office of Polar Programs,
U.S. National Science Foundation-Physics Division,
U.S. National Science Foundation-EPSCoR,
U.S. National Science Foundation-Office of Advanced Cyberinfrastructure,
Wisconsin Alumni Research Foundation,
Center for High Throughput Computing (CHTC) at the University of Wisconsin{\textendash}Madison,
Open Science Grid (OSG),
Partnership to Advance Throughput Computing (PATh),
Advanced Cyberinfrastructure Coordination Ecosystem: Services {\&} Support (ACCESS),
Frontera and Ranch computing project at the Texas Advanced Computing Center,
U.S. Department of Energy-National Energy Research Scientific Computing Center,
Particle astrophysics research computing center at the University of Maryland,
Institute for Cyber-Enabled Research at Michigan State University,
Astroparticle physics computational facility at Marquette University,
NVIDIA Corporation,
and Google Cloud Platform;
Belgium {\textendash} Funds for Scientific Research (FRS-FNRS and FWO),
FWO Odysseus and Big Science programmes,
and Belgian Federal Science Policy Office (Belspo);
Germany {\textendash} Bundesministerium f{\"u}r Forschung, Technologie und Raumfahrt (BMFTR),
Deutsche Forschungsgemeinschaft (DFG),
Helmholtz Alliance for Astroparticle Physics (HAP),
Initiative and Networking Fund of the Helmholtz Association,
Deutsches Elektronen Synchrotron (DESY),
and High Performance Computing cluster of the RWTH Aachen;
Sweden {\textendash} Swedish Research Council,
Swedish Polar Research Secretariat,
Swedish National Infrastructure for Computing (SNIC),
and Knut and Alice Wallenberg Foundation;
European Union {\textendash} EGI Advanced Computing for research;
Australia {\textendash} Australian Research Council;
Canada {\textendash} Natural Sciences and Engineering Research Council of Canada,
Calcul Qu{\'e}bec, Compute Ontario, Canada Foundation for Innovation, WestGrid, and Digital Research Alliance of Canada;
Denmark {\textendash} Villum Fonden, Carlsberg Foundation, and European Commission;
New Zealand {\textendash} Marsden Fund;
Japan {\textendash} Japan Society for Promotion of Science (JSPS)
and Institute for Global Prominent Research (IGPR) of Chiba University;
Korea {\textendash} National Research Foundation of Korea (NRF);
Switzerland {\textendash} Swiss National Science Foundation (SNSF).

\end{document}